\documentclass[%
 reprint,longbibliography,preprintnumbers,
nofootinbib,
 amsmath,amssymb,
 aps,
prl,
]{revtex4-1}
\pdfoutput=1
\usepackage[utf8]{inputenc}
\usepackage{flushend}
\usepackage{dcolumn}% Align table columns on decimal point
\usepackage{bm}% bold math
\usepackage{balance}
\usepackage[normalem]{ulem}

\usepackage[colorlinks = true,
            linkcolor = blue,
            urlcolor  = blue,
            citecolor =green,
            anchorcolor = blue]{hyperref}
\usepackage{verbatim}
\usepackage{color,ulem}
\usepackage[english]{babel}

\usepackage[utf8]{inputenc}
\input Starburst.fd
\newcommand*\initfamily{\usefont{U}{Starburst}{xl}{n}}\initfamily

\newcommand{\beq}{\begin{eqnarray}}
\newcommand{\eeq}{\end{eqnarray}}
\usepackage{amsmath}
\usepackage{tikz}
\usetikzlibrary{decorations.pathmorphing}
\usetikzlibrary{shapes.misc}
\tikzset{cross/.style={cross out, draw=black, minimum size=8*(#1-\pgflinewidth), inner sep=0pt, outer sep=0pt},
cross/.default={1pt}}
\usetikzlibrary{patterns,math}
\begin{document}

\title{Random close packing of binary hard spheres favors \\ the stability of neutron-rich atomic nuclei}

\author{\textbf{Carmine Anzivino$^{1}$}}
\author{\textbf{Vinay Vaibhav$^{1}$}}
\author{\textbf{Alessio Zaccone$^{1,2}$}}%
 \email{alessio.zaccone@unimi.it}

 \vspace{1cm}
 
\affiliation{$^{1}$Department of Physics ``A. Pontremoli'', University of Milan, via Celoria 16,
20133 Milan, Italy.}

\affiliation{$^{2}$Institute of Theoretical Physics, University of G\"ottingen, Friedrich-Hund-Platz 1, 37077 G\"ottingen, Germany.}

\begin{abstract}
In spite of the success of the Bethe-Weizsäcker mass formula in its modern numerical and predictive implementations, the common-knowledge principle that it is electrostatics which, ultimately, favors neutron-rich nuclei still presents unclear aspects.
For example, while it is true that the Coulomb interaction promotes the tendency towards neutron-rich nuclei, the opposite effects of Majorana exchange forces and Pauli exclusion are known to counteract this tendency.
We show that a recent analytical progress in the mathematical description of random close packing of spheres with different sizes provides a missing contribution to the theoretical description of the $Z$ versus $N$ slope in the nuclides chart. In particular, the theory suggests, on geometric grounds and with a physically-reasoned assumption that the excluded-volume size of neutrons is 20\% larger than that of protons, that the most stable nuclei are those with ratio $Z/N\approx 0.75$. This new ``geometric'' random-packing contribution to the semi-empirical mass formula may be the missing aspect of nuclear structure that tilts the balance towards neutron-rich nuclei in the Segrè stability chart.

\end{abstract} 

\maketitle
Meitner and Frisch \cite{Meitner1939}, in their famous 1939 explanation of nuclear fission, were among the first to emphasize the classical analogy between nucleons in atomic nuclei and the random close packing of hard spheres. This analogy lies at the heart of the liquid drop model of nuclei \cite{Weizsäcker1935,Bethe1936,Frenkel}.

The main success of the liquid drop model is the Bethe-Weizsäcker semi-empirical mass formula, which gives the binding energy $B$ of the nucleus (from which the nuclear mass is estimated via the relativistic mass defect) as \cite{Obertelli2021}:
\begin{equation}
    B=a_V\,A -a_S\,A^{2/3}-a_C\,\frac{Z(Z-1)}{A^{1/3}}-a_A \frac{(N-Z)^2}{A}+\frac{\delta} {A^{1/2}}\label{Weiz}
\end{equation}
where $A=Z+N$. The first term on the right hand side is the so-called volume term due to nucleon-nucleon attractive interaction energy, the second term is the surface term, the third one is the Coulomb repulsion term due to the protons, the fourth term is the asymmetry term, and the last one is the pairing term. The respective parameters ($a_V$, $a_S$, $a_C$, $a_A$, $\delta$) are empirical coefficients, the values of which are determined by fitting experimental data from mass spectroscopy \cite{Mottelson}. Of these 5 fitting parameters, perhaps only the Coulomb term can be precisely determined from theory. More sophisticated versions of Eq. \eqref{Weiz} exist \cite{Myers}, as nuclear structure theory has developed into a field of
systematic calculations and predictions based on ab initio few- and
many-body calculations, reaching nuclei as heavy as $^{208}$Pb, as well as
based on global energy density functional calculations of the nuclear stability
chart.
With these auxiliary microscopic calculation methods, Eq. \eqref{Weiz} can be used to provide an excellent fitting to both the binding energy curve of all nuclides and to the Segrè chart of stability of nuclides (where for each nuclide the proton number $Z$ is plotted as a function of the neutron number $N$) \cite{Segre}. 

In spite of this, in an attempt to extract general physical principles, it is still interesting to study Eq. \eqref{Weiz} from the point of view of how the interplay between different physical phenomena leads Eq. \eqref{Weiz} to capture the actual observed stability trend, i.e. $Z \approx 0.75 N$ \cite{Marmier}.

The standard qualitative understanding of the Segrè nuclear stability chart based on Eq. \eqref{Weiz} is that the Coulomb repulsion, for large nuclei, grows as $\sim Z^2$, whereas the attractive energy due to the strong force increases only as $\sim Z$ (due to its short-range character). This fact justifies the tendency of larger nuclei to have more neutrons than protons, $N > Z$, in order to compensate the unfavourable growth of the Coulomb energy \cite{Segre}. However, this energetic gain is, in its own turn, compensated or reduced by the asymmetry between neutrons and protons, which corresponds to the energetically unfavourable asymmetry term, i.e. the fourth term in Eq. \eqref{Weiz}. In a nutshell, this term arises due to the Pauli exclusion principle, because having more neutrons than protons ($N > Z)$ implies that the neutrons have to be stacked up into higher energy levels. More specifically, there are two contributions from the asymmetry (neutron excess). The first is an increase in the total potential energy, which can be qualitatively estimated by considering that the nuclear wavefunction is anti-symmetric with respect to exchanging two identical particles in different space states. The energy cost of interchanging protons with neutrons can be approximately evaluated by making use of the Majorana space-exchange forces, leading to a contribution to the binding of the form $\sim (N-Z)^2/A$ \cite{Blatt,Preston}.
The corresponding increase in kinetic energy can be evaluated, in a rudimentary way, using a Fermi gas model and also contributes a term of the form $\sim (N-Z)^2/A$ \cite{Blatt,Preston}. Combining the two contributions with a common prefactor leads to the fourth term on the r.h.s. of Eq. \eqref{Weiz}. The prefactor $a_c$ is necessarily empirical since it is prohibitively difficult to estimate it from theory for a large number of nuclides (and it would necessarily be inaccurate given the approximate character of the theory of exchange forces in nuclei) \cite{Obertelli2021}. 

We are thus left with the qualitative conclusion that the neutron excess in stable nuclei is dictated by a subtle balance between the electrostatic energy (Coulomb term) and the corresponding asymmetry term in Eq. \eqref{Weiz}. It is, therefore, this subtle balance, that cannot be theoretically predicted, which ultimately determines the stability of nuclei. This balance is subtle because, if the two competing effects were roughly equal, then Eq. \eqref{Weiz} would predict that the most stable nuclides are those with $N=Z$, at odds with experimental observations of neutron excess $Z/N>1$ in the Segrè chart.

Here, our focus is not to make accurate quantitative predictions of nuclear stability, which is a task that is best achieved using the current semi-empirical numerical packages \cite{Colo}. Rather, we want to address an aspect of nuclear stability that has been neglected so far. Common knowledge identifies the main source of isospin breaking in nuclei with the long-range Coulomb interaction, with possibly a subsidiary role played by quark masses. It cannot be denied, however, that the effect of the Coulomb interaction may be compensated or cancelled out by the opposite effect of the exchange forces and by the Pauli principle, as discussed above. Given the presence of empirical parameters, state-of-art numerical calculations cannot provide transparent insights into this question.
While keeping the analysis at a fundamental level, we want to show that there is a further structural aspect, neglected in the literature so far, which effectively helps the Coulomb repulsion to win the balance towards favoring neutron-rich nuclei. This effect is purely structural and has to do with the maximal packing density that a mixture of spheres with two different sizes can achieve. This aspect is not in contradiction or in opposition with all the other (mostly quantum-mechanical) aspects of nuclear structure recalled above, and it would represent an additional, new contribution to the semi-empirical mass formula, acting in synergy with the known ones to produce the observed stability trend.

Recently, the original analogy between random close packing (RCP) \cite{Torquato_review} of hard spheres and the density of nuclear matter has been used in estimates of nucleon density in the context of neutron stars \cite{Kaiser_neutron} and of nucleon sizes \cite{kaiser2024sizes}. 
These recent estimates took advantage of an analytical solution to the RCP problem proposed in 2022 \cite{Zaccone_RCP}. This solution was found by considering that RCP should be a point in the ensemble of jammed states of hard spheres where mechanical stability vanishes or, equivalently, where the sphere packing is isostatic, with coordination number $z = 6$ in
3d \cite{Scossa,zaccone2023}. Since the “minimally coordinated” mechanically-stable sphere packings for each density are those closest in
structure to liquid (non mechanically-stable) states, one can model the coordination number by well-known analytical theories
for the equation of state (EOS) of a hard-sphere liquid. The underlying principle is the analogy between the crowding of atoms in a liquid and jammed states of hard spheres, that was already pointed out by Kamien and Liu in a seminal paper \cite{Kamien}.

The analytical theory for RCP of monodisperse spheres \cite{Zaccone_RCP} has opened up the possibility to have an analytical prediction also for RCP of hard spheres of different size, i.e. with a size distribution (spheres with polydispersity in diameter) or mixtures of spheres having two distinct diameters (binary mixtures). In both cases, the theory produces analytical predictions in agreement with numerical simulations as shown in \cite{Anzivino_JCP}, without adjustable parameters.

In the following, we apply the theory of RCP for binary mixtures to atomic nuclei and show that it is able to predict the correct stability trend of atomic nuclei. 
We start from the derivation of the RCP for binary mixtures of hard spheres with two different diameters, $\sigma_1$ and $\sigma_2$, representing, respectively, the diameter of neutrons and protons. 
The full details of the derivation can be found in previous work \cite{Anzivino_JCP}, here we limit ourselves to a brief recap. 

We denote as  $\sigma_{ij} = \frac{1}{2} (\sigma_{i} + \sigma_{j})$ the contact distance between a sphere of species $i$ and a sphere of species $j,$ where $\sigma_{ii} \equiv \sigma_{i}$ is the diameter of a sphere of species $i.$ Here, $i,j = N,Z$.
We denote the number fraction of species $i$ as $x_{i}= \rho_{i}/\rho,$ where $\rho$ is the number density of the nucleus, while $\rho_{i}$ is the number density of species $i.$ 
Then, we define $\left\langle \sigma^{n} \right\rangle \equiv \sum_{i=1}^{m} x_{i} \sigma_{i}^{n}$ such that the total volume fraction (volume occupied by all the spheres over the total volume of the system) is given by $\phi = \pi \rho \left\langle \sigma^{3} \right\rangle /6.$ 
As is standard in liquid state theory \cite{Hansen}, the coordination number or mean contact number between a sphere and its nearest neighbours is given by:
\begin{equation}
z_{ij} = 4 \pi \rho \int_{0}^{\sigma_{ij}^+} dr r^2 g_{ij}(r).
\end{equation}
where $g_{ij}(r)$ is the standard pair correlation function defined in elementary statistical mechanics \cite{Chandler} and $r$ is the radial separation coordinate. In the above expression, $\sigma_{ij}^+ = \sigma_{ij} + \epsilon$ where $\epsilon$ is in an infinitesimally small positive number. The only part of $g_{ij}(r)$ which contributes to the coordination number is the contact value $g_{ij,c}$, defined as \cite{Anzivino_JCP}: $g_{ij,c}(\sigma_{ij}; \phi) \equiv \frac{\sigma_{ij}}{\langle \sigma \rangle} g_0(\langle \sigma \rangle) g_{ij}(\sigma_{ij}; \phi),$ where $g_0$ is a dimensional constant (with dimension $1/length$) to be determined based on a limiting reference state (usually taken as the closest packing of spheres in 3D provided by the face centered cubic, fcc, lattice \cite{Hales1}).
One thus obtains the stability condition \cite{Anzivino_JCP}:
\begin{align}
\langle z \rangle  &=6= 24 \phi \frac{g_0}{\langle\sigma\rangle} \sum\limits_{i,j}^{m} x_i x_j \frac{\sigma_{ij}^3}{\langle \sigma^3 \rangle} g_{ij}(\sigma_{ij}; \phi),\label{key}
\end{align}
where $m=N,Z$. 
This expression can be evaluated analytically by using the compressibility factor $\mathcal{Z}^{(m)}$ (and therefore the species-averaged pair correlation function at contact appearing in the virial theorem, $g^{(m)}_{\textrm{eq}}$) of equilibrium binary hard spheres derived by Lebowitz 
 \cite{lebowitz}
\begin{equation} \label{Z_mixture}
 g^{(m)}_{\textrm{eq}}(\sigma;\phi) \equiv \frac{\mathcal{Z}^{(m)} (\phi) - 1}{4 \phi} =  \sum_{i=1}^{m} \sum_{j=1}^{m} x_{i} x_{j} \frac{\sigma_{ij}^{3}}{\left\langle \sigma^{3} \right\rangle} g_{ij} (\sigma_{ij}; \phi),
\end{equation}
where the approximation of taking the equilibrium estimate for the contact value of the pair correlation function has been justified extensively on the basis of numerical simulations in \cite{Anzivino_JCP}. We assume the compressibility factor $\mathcal{Z}^{(m)} (\phi)$ to be \cite{santos_yuste_deharo_1999}
\begin{equation} \label{Zsantos}
\begin{aligned}
\mathcal{Z}^{(m)} & (\phi) = 1 + \big[ \mathcal{Z}_{PY} (\phi) - 1 \big] \frac{\left\langle \sigma^{2} \right\rangle}{2 \left\langle \sigma^{3} \right\rangle^{2}} \big( \left\langle \sigma^{2} \right\rangle^{2} + \left\langle \sigma \right\rangle \left\langle \sigma^{3} \right\rangle \big) \\
&+ \frac{\phi}{(1-\phi)} \bigg[ 1 - \frac{\left\langle \sigma^{2} \right\rangle}{\left\langle \sigma^{3} \right\rangle^{2}}  \big( 2 \left\langle \sigma^{2} \right\rangle^{2} - \left\langle \sigma \right\rangle \left\langle \sigma^{3} \right\rangle \big) \bigg],
\end{aligned}
\end{equation}
% \begin{align}
where $\mathcal{Z}_{PY} (\phi)$ is the Percus-Yevick compressibility factor \cite{Hansen}. Using this expression for $\mathcal{Z}^{(m)}  (\phi)$ in Eq. \eqref{Z_mixture}, and the latter in Eq. \eqref{key}, one can solve Eq. \eqref{key} for $\phi \equiv \phi_{RCP}$. This is the RCP volume fraction, which is a function of the size ratio between the spheres, $\sigma_2/\sigma_1$, and also a function of the composition of the mixtures, i.e. the ratio of big spheres to small spheres present in the systems, in our case $N/Z$ or, equivalently $N/A$, by taking into account that $A=N+Z$.

We now consider the following estimates for the proton radius and for the neutron radius. For the proton, the internationally accepted value of its radius is $\sigma_2/2 \equiv r_Z= 0.84$ fm \cite{NIST}. For the neutron, there is no such clear cut value, given the well known difficulties in experimentally determining the size of electrically neutral particles. It would not make sense to use the root mean square charge radius here since it is related to the neutron-electron scattering length, hence to how neutrons interact with electrons, which is certainly irrelevant and misleading for our purpose of modelling excluded-volume interactions between nucleons in a densely packed nucleus. 
A more meaningful choice for our purpose of modelling excluded-volume between nucleons is to take the size at which the neutron charge density profile goes to zero, which is at $\sigma_1 \equiv r_N \approx 1.0$ fm, according to the latest literature \cite{Atac2021}. This is of course, a hypothesis of the current calculation, which should be granted the benefit of doubt, but it does seem to be the most reasonable one given the alternatives at hand and is not disproved by the available empirical evidence (here one should also be aware of the current unsettled debate on the sizes of the nucleons, especially given the ambiguities which plague the precise determination of charge density distributions of nucleons \cite{Jaffe,Epelbaum_hadrons}).

Based on these considerations, we thus obtain the following estimate for the size ratio of neutrons and protons: $r_Z/r_N = \sigma_2/\sigma_1=0.84$.

By analytically solving Eq. \eqref{key} for $\phi$ with $\sigma_2/\sigma_1=0.84$, we thus obtain the RCP volume fractions plotted in Fig. 1(a) as a function of the ratio of small spheres over the total (in our case, the ratio of protons to the total number of nucleons in the nucleus). We also show, in Fig. 1(b), an artistic rendering of the RCP structure with protons and neutrons having the sizes and composition at which the theory predicts the largest packing density (i.e. the maximum of the curve in Fig. 1(a)).
For illustrative purposes we chose a simple spherical shape, but it is important to point out that the predicted RCP volume fraction is independent of the nucleus shape. That is, the model predicts the same maximal RCP volume fraction value (for $Z/A=0.43$) also for different shapes, e.g. ellipsoidal or elongated nuclei, and deformed nuclei.

\begin{figure}[h]
\centering
\includegraphics[width=0.6\linewidth]{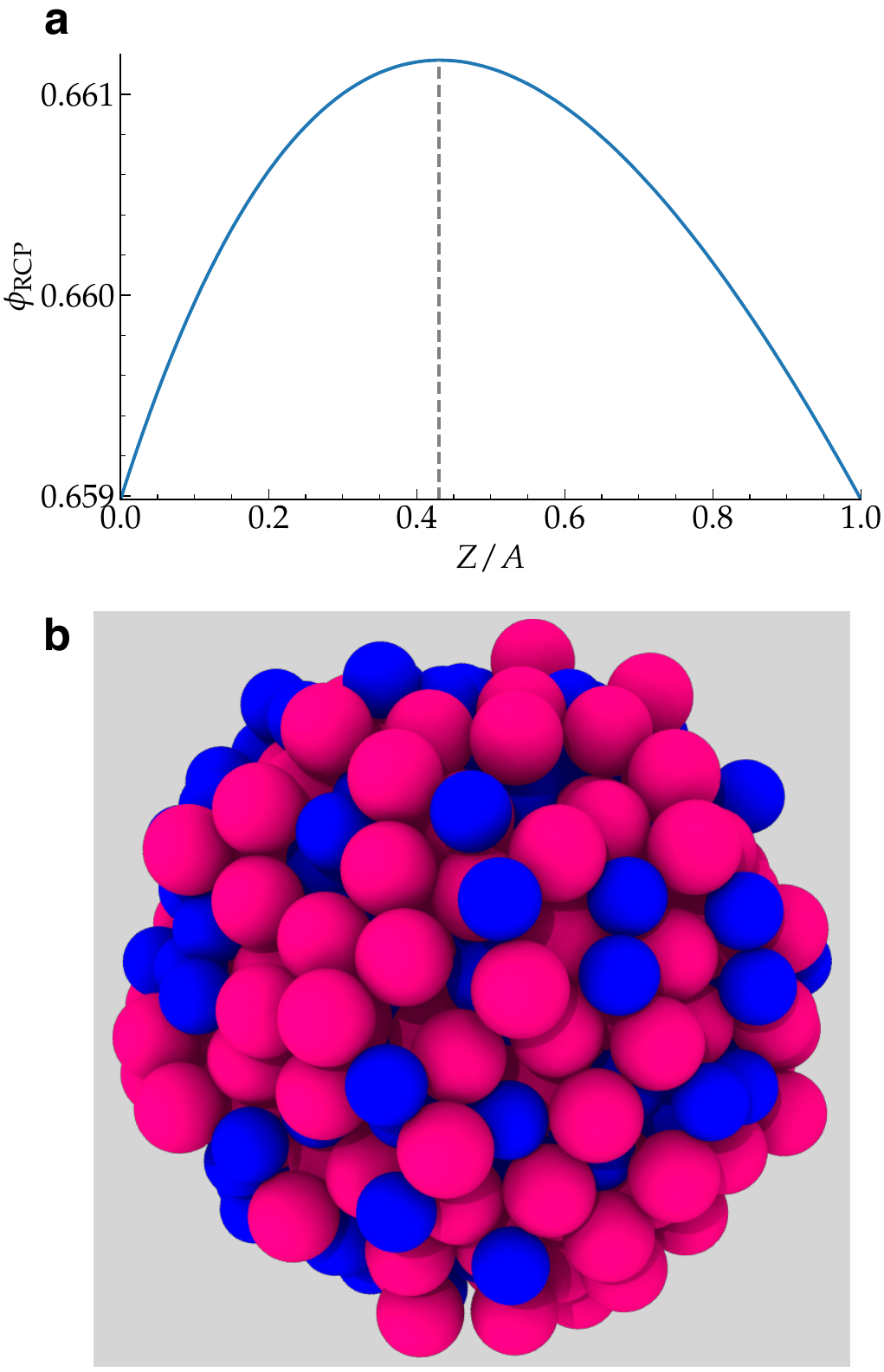}
\caption{Panel (a) shows the random close packing (RCP) volume fractions computed, for binary hard sphere mixtures where the small spheres have the radius of the proton and the big spheres have the radius of the neutron, based on Eqs. \eqref{key}-\eqref{Zsantos}. The theory predicts a maximum RCP density corresponding to the composition $Z/A = 0.43$ (dashed vertical line). Panel (b) shows a schematic rendering of an RCP structure corresponding to the largest packing fraction predicted by the theory (indicated by a dashed vertical line in panel (a)), i.e. with $Z/A = 0.43$ and with size ratio $r_Z/r_N = \sigma_2/\sigma_1=0.84$. Red spheres are neutrons and blue spheres are protons. For simplicity, no effects like neutron skin or $\alpha$-particle clustering are taken into account in the picture. The spherical shape was chosen for illustrative purposes, and the predicted RCP volume fraction would be exactly the same also for different nuclear shapes (e.g. ellipsoidal, elongated etc).}
\label{fig1}
\end{figure}

By computing the maximum of the curve in Fig. 1(a), we thus predict that the maximum random close packing nuclear density is achieved for
\begin{equation}
Z/A = 0.43
\end{equation}
which, in turn, directly implies
\begin{equation}
Z/N = 0.75
\end{equation}
which, remarkably, coincides with the most quoted empirical trend value for the slope in the nuclides stability (Segrè) chart \cite{Preston}.

Using the most recent data for the composition of stable nuclides \cite{IAEA}, the predicted trend is shown in Fig. 2 (orange line).

\begin{figure}[h]
\centering
\includegraphics[width=\linewidth]{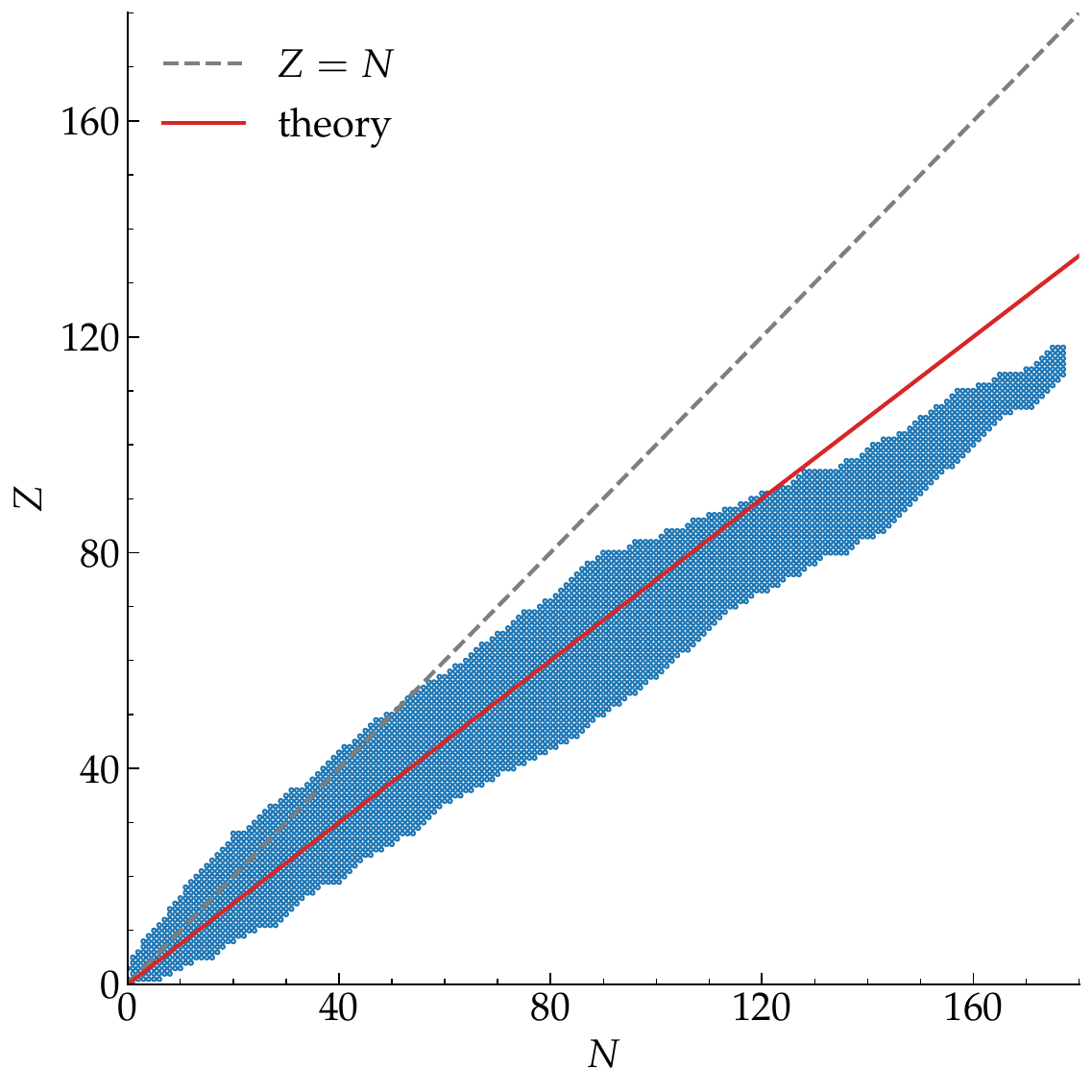}
\caption{The Segrè chart with stable nuclides plotted in terms of their proton number $Z$ versus neutron number $N$. The blue symbols are the most recent experimental data provided by the International Atomic Energy Agency (IAEA) \cite{IAEA}. The grey line is the symmetry line. The red line is the prediction of this paper, based on Eq. \eqref{key}.}
\label{fig2}
\end{figure}

The fact that the maximum packing density is achieved when there is an excess of the large spheres over the small spheres (the former are 57\% of the total) is due to purely geometric reasons. This is because, in general, space is filled more efficiently if there is a majority of larger spheres with smaller spheres filling up the empty gaps between the former.

From a physical point of view, the above theory suggests that the neutron-excess observed in stable large nuclei, besides the subtle balance between electrostatics and Pauli-exclusion asymmetry energy, can be understood in terms of the tendency of the nuclei to minimize their energy by finding the configuration with the largest volume fraction occupied by the nucleons, i.e. with the highest density of nucleons. This can be further understood as follows.
Due to the short-range character of the strong force, the central attractive interaction between nucleons becomes negligible at a nucleon-nucleon separation of about 3 fm \cite{Epelbaum1,Epelbaum2}. 
This implies that mostly nearest-neighbour and next nearest-neighbour interactions contribute to the binding energy, which is, after all, textbook knowledge \cite{Obertelli2021}. If the nearest-neighbour coordination number is $z$, the total number of pairwise nucleon-nucleon interactions in the nucleus is $z A/2$ (which justifies the first term in Eq. \eqref{Weiz} being $\propto A$). If each nucleon has, on average, $z$ nearest neighbours and $z_{nn}$ next nearest neighbours, the total number of interactions contributing to the binding energy is $z_{tot}A/2=(z+z_{nn})A/2$. In the RCP of binary spheres, the total average $z$ is always 6 regardless of the composition i.e. regardless of $Z/A$. However, the mean number of next nearest neighbours is, instead, larger the larger the volume fraction $\phi_{RCP}$. This can be shown more quantitatively by recalling that $z_{tot}$ is given by:
\begin{equation}
z_{tot} = 24 \phi_{RCP} \int_{1}^{x_c} dx x^2 g_{tot}(x).
\end{equation}
where $x \equiv r/\sigma$ with $\sigma$ the average nucleon diameter, and $x_c$ is the range of the nuclear strong force.
Hence, since the main contribution to the binding energy $B$ is given by $a_V A$ which therefore is proportional to $z_{tot} A/2 \propto \phi_{RCP}$, it is evident that the most stable nuclei will be those with the largest $\phi_{RCP}$ value. According to the above theory, those are, indeed, the nuclei with $Z/N = 0.75$.

In summary, we have presented a contribution to the theory of the stability of atomic nuclei in terms of random close packing of binary hard spheres. The theory shows that the binding energy is maximized by the largest nucleons packing density, which is estimated using the statistical mechanics theory of random close packing (RCP) of bi-disperse hard spheres \cite{Anzivino_JCP}. Assuming reasonable values for the proton radius and for the neutron radius, the theory predicts that the largest density is achieved when $Z/N = 0.75$, the most quoted slope value in the literature \cite{Preston}. 

This result opens up new possibilities in nuclear structure and in low energy nuclear physics. Most importantly, it provides a new rational and predictive guideline in the experimental search for new heavy nuclei and the stability thereof \cite{Block2010,Morjean}. It may also have broader implications for nuclear reactions such as fission \cite{Wilson2021} and fusion-fission \cite{Eccles}, for the equation of state of nuclear matter \cite{EOS,EOS2}, for the viscosity and hydrodynamics of nuclear matter \cite{Noronha_2022,Zaccone2024}, and nucleosynthesis \cite{Sun2024}. In future work, the proposed ideas may also be extended to deformed warm nuclei to describe dipole oscillations \cite{Bracco}.

\subsection{Acknowledgments} 
We are indebted to Prof. Gianluca Colò (University of Milan) for suggesting Ref. \cite{IAEA} and to Prof. Angela Bracco (University of Milan and The Enrico Fermi Research Center Rome) for input and discussion. A.Z. and V.V. gratefully acknowledge funding from the European Union through Horizon Europe ERC Grant number: 101043968 ``Multimech''. A.Z. gratefully acknowledges the Nieders{\"a}chsische Akademie der Wissenschaften zu G{\"o}ttingen in the frame of the Gauss Professorship program and funding from US Army Research Office through contract nr. W911NF-22-2-0256.
%I gratefully acknowledge funding from the European Union through Horizon Europe ERC Grant number: 101043968 ``Multimech'', and from US Army Research Office through contract nr. W911NF-22-2-0256. 

\bibliographystyle{apsrev4-1}

\bibliography{refs}

\end{document}